\documentclass[10pt,twocolumn,letterpaper]{article}
\usepackage[pagenumbers]{FORMATARXIV} 
\usepackage{graphicx}
\usepackage{amsmath}
\usepackage{amssymb}
\usepackage{booktabs}
\usepackage{multirow} 
\usepackage[table]{xcolor}
\usepackage{tabularx}

\definecolor{darkyellow}{rgb}{1, 0.753, 0}
\usepackage[pagebackref,breaklinks,colorlinks]{hyperref}
\usepackage[capitalize]{cleveref}
\crefname{section}{Sec.}{Secs.}
\Crefname{section}{Section}{Sections}
\Crefname{table}{Table}{Tables}
\crefname{table}{Tab.}{Tabs.}

\usepackage{todonotes}

\usepackage{xcolor}
\definecolor{lightblue}{RGB}{107, 174, 214}
\begin{document}

\title{AC-IND: Sparse CT reconstruction based on attenuation coefficient estimation and implicit neural  distribution}
\author{
Wangduo Xie$^{1}$, Richard Schoonhoven$^{2}$, Tristan van Leeuwen$^2$, 
Matthew B. Blaschko$^1$ \\
$^1$ {ESAT-PSI, KU Leuven, 3001 Leuven, Belgium},\\
$^2$ {Centrum Wiskunde \& Informatica, 1098 XG Amsterdam, Netherlands} \\
\vspace{-1em}
\\{\tt\small wangduo.xie@kuleuven.be, \{richard.schoonhoven, t.van.leeuwen\}@cwi.nl,}
\\{\tt\small matthew.blaschko@esat.kuleuven.be}
}
\maketitle

\begin{abstract}
   Computed tomography (CT) reconstruction plays a crucial role in industrial nondestructive testing and medical diagnosis. Sparse view CT reconstruction aims to reconstruct high-quality CT images while only using a small number of projections, which helps to improve the detection speed of industrial assembly lines and is also meaningful for reducing radiation in medical scenarios. Sparse CT reconstruction methods based on implicit neural representations (INRs) have recently shown promising performance, but still produce artifacts because of the difficulty of obtaining useful prior information. In this work, we incorporate a powerful prior: the total number of material categories of objects. To utilize the prior, we design \textbf{AC-IND}, a self-supervised method based on \textbf{A}ttenuation \textbf{C}oefficient Estimation and \textbf{I}mplicit \textbf{N}eural \textbf{D}istribution. 
   Specifically, our method first transforms the traditional INR from scalar mapping to probability distribution mapping. Then we design a compact attenuation coefficient estimator initialized with values from a rough reconstruction and fast segmentation.  Finally, our algorithm finishes the CT reconstruction by jointly optimizing the estimator and the generated distribution. Through experiments, we find that our method not only outperforms the comparative methods in sparse CT reconstruction but also can automatically generate semantic segmentation maps.
   
\end{abstract}

\section{Introduction}
\label{sec:intro}
Computed tomography (CT) is a non-destructive testing technology that can image the inside of an object in a non-invasive way, which helps to measure structure and find defects. It is also widely used in medical imaging for disease diagnosis. Therefore, improving the accuracy of CT reconstruction is a classic problem of concern.

\begin{figure}[ht]
    \centering
    \includegraphics[width=8cm]{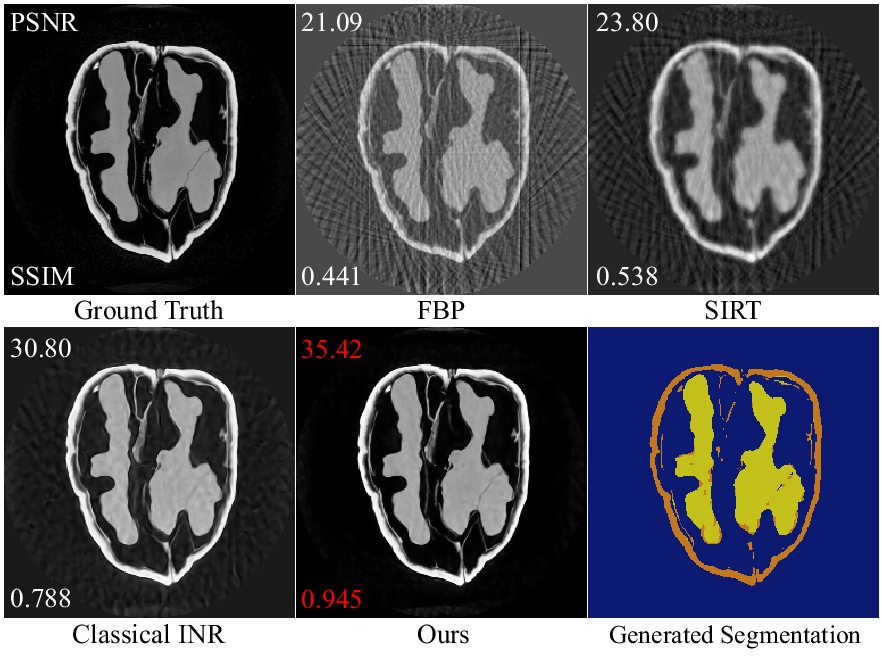} 
   
    \caption{Performance comparison on the reconstruction image and segmentation image \textbf{automatically generated} by our method.}
    \label{fig:begin_figure}
\end{figure}

Usually, dense projection leads to more accurate reconstruction compared to sparse projection. However, obtaining dense projections is time-consuming, posing a significant challenge for industrial factories that handle large quantities of products. In addition, dense projection increases the total radiation dose received by patients in medical applications. To this end, in order to achieve better sparse reconstruction, a large number of deep learning-based methods \cite{jin2017deep, zhang2018sparse, wu2021drone,he2020radon,han2016deep} have been proposed in recent years. However, these methods belong to supervised learning and require many projection-image pairs to complete the learning. This easily leads to difficulty in generalization because of the domain gap between the training source and the testing source. Although a supervised generative model\cite{ying2019x2ct, songsolving} can slightly alleviate the problem, it often brings artifacts, which limits applicability.

Implicit Neural Representations (INRs) have recently shown impressive results for sparse CT reconstruction\cite{tancik2020fourfeat,sun2021coil}. However, from a Bayesian perspective, INRs  merely provide a maximum likelihood estimate of the object and fail to incorporate a prior. While INRs avoid the domain gap issues associated with supervised learning, the large solution space can result in reconstructions that, despite having minimal projection loss, may not accurately reflect the object's natural appearance. Previous works \cite{shen2022nerp, gu2023generalizable} attempted to embed prior into INRs by compressing external information through supervised learning, and then assisting INRs in reconstructing new objects. Although these approaches have yielded good results, the reliance on supervised learning leads to a reconstruction degradation when the current object undergoes a large change. This limitation motivates our paper, which aims to embed prior information into INRs without being affected by distribution shift.

In addition, the backbone of INRs for CT reconstruction is usually an MLP with a trigonometric activation function. According to the black box nature of MLPs, the physical meaning is only known for the last neuron, which represents the reconstructed linear attenuation coefficient of the object, while the roles of the other neurons in the MLP for reconstruction remain unclear. Therefore, increasing interpretability in CT reconstruction also gave rise to this work.

To effectively utilize prior information without introducing domain gaps, and to increase the method's interpretability, this paper proposes a self-supervised CT reconstruction method based on \textbf{A}ttenuation \textbf{C}oefficient estimation and  \textbf{I}mplicit \textbf{N}eural \textbf{D}istribution (\textbf{AC-IND}). Specifically, our algorithm only requires prior knowledge of the number of materials within the object to be reconstructed. The prior information can be obtained without relying on a learning algorithm based on external data. It's generally a basic attribute of industrial objects and is easily accessible.  

However, the traditional INR, being an MLP with only a single neuron in the final layer, is unable to directly incorporate the number of material categories. Therefore, this work no longer maps the discrete coordinates to a scalar like the traditional INR. Instead, each coordinate is mapped to a distribution, which describes the degree of each coordinate belonging to a different material category. To encourage the distribution to converge toward its true form during training, AC-IND first produces a modulated distribution to make it close to a steep unimodal distribution. Then, it employs rapid reconstruction methods along with Multi-Otsu Thresholding method\cite{liao2001fast}  to obtain a rough estimation of the average attenuation coefficient (AC) within each material. Afterwards, a compact differentiable AC estimator is designed and initialized with the rough estimation. Finally, the integral of the AC estimator's output under the produced distribution is the output value of our algorithm. Under the indirect supervision of projection data, AC-IND progressively reconstructs an accurate CT image through the joint training of the modified INR and AC estimator.

The experiments demonstrate that AC-IND achieves superior reconstruction accuracy compared to the classic INR, at sparse view counts of 20, 40, and 60. Additionally, we found that the produced distribution of our method exhibits clear physical meaning. Specifically, by choosing the most likely material based on the distribution's peak, we discovered that our method learns the segmentation map in an unsupervised manner. Furthermore, the segmentation map,  continuously learned AC, and reconstruction result exhibit synchronized improvement during training.  

The synchronized improvement suggests that generating more accurate AC values from the start will enhance reconstruction. Building on this insight, we developed an enhanced algorithm named AC-IND\textsuperscript{+}. In AC-IND\textsuperscript{+}, the AC estimator is initialized with the mean values of different segmentation areas, derived from reconstruction results of AC-IND and the Multi-Otsu Thresholding method. Experiments indicate that improved initialization of AC estimator leads to enhanced reconstruction and segmentation.

In summary, our contributions are as follows:

1. We propose a sparse CT reconstruction algorithm named AC-IND, based on attenuation coefficient estimation and implicit neural distribution. Additionally, we introduce an enhanced version named AC-IND\textsuperscript{+} with better initialization of AC estimator.

2. Our method generates semantic segmentation images during reconstruction that the comparative algorithm is unable to produce, creating an opportunity to merge CT reconstruction and unsupervised semantic segmentation tasks.

3. We conduct a dynamic analysis of the distribution generated by our method and the continuous evolution of AC estimator's output during training. The analysis reveals that as reconstruction performance improves, the automatically generated segmentation map becomes more accurate, and the value produced by the AC estimator eventually approaches the ground truth of the AC.

4. Experiments demonstrate that AC-IND outperforms the comparative approach in CT reconstruction. The performance of AC-IND\textsuperscript{+} highlights the improved initialization of AC estimator can lead to enhanced reconstruction.

\section{Related Work}
\label{sec:Related Work}

\subsection{Classical CT reconstruction}
Classical CT reconstruction methods can be broadly categorized into analytical and iterative approaches. A well known analytical method is filtered back projection (FBP) \cite{herman2009fundamentals}, which is based on the Fourier slice theorem. When the projection data is dense, FBP can yield good results; however, it performs poorly with sparse views. Examples of iterative methods
are represented by the simultaneous iterative reconstruction technique (SIRT), simultaneous algebraic reconstruction technique (SART) \cite{andersen1984simultaneous}, and other algorithms \cite{gordon1970algebraic, sauer1993local, fessler1994penalized, yu2010fast}  based on iterative optimization. However, these methods are time-consuming and struggle to reconstruct high-precision images.

\subsection{CT reconstruction with INR}
Implicit neural representation (INR) completes the reconstruction process by using a neural network to map vectors in coordinate space to real physical quantity under the supervision of the observed measurement data. In works \cite{tancik2020fourfeat,cai2024batch}, its interpretability and dynamics analysis were developed using neural tangent kernels \cite{jacot2018neural} and random Fourier features \cite{rahimi2007random}. Accordingly, It has shown good performance in new perspective image synthesis \cite{mildenhall2021nerf} and 3D reconstruction \cite{rematas2022urban} for natural scenes.
Recently, it shown good performance for CT reconstruction \cite{zha2022naf,wu2023unsupervised,saragadam2023wire,sun2021coil,zang2021intratomo,ruckert2022neat,wu2023self,shi2024implicit}. However, when optimizing INR, this type of method does not include other prior information about the object except the basic statistical laws of the image, such as TV regularization \cite{chambolle2004algorithm}. Due to the presence of spectral bias \cite{rahaman2019spectral, yuce2022structured}, INR struggles to fit high-frequency information and lead to blurred edges and artifacts. Therefore, incorporating prior knowledge is expected to enhance INR's performance. 

\subsection{CT reconstruction and external prior} 
To utilize priors, \cite{gu2023generalizable,song2023piner,shen2022nerp,lin2023learning, tancik2021learned,kazerouni2024incode}  integrate external datasets into INR through supervised methods. However, there is a domain gap between the external data and the data we need to reconstruct, which deteriorates the performance of the machine learning algorithm\cite{xie2023dense,zhang2021adaptive}. Therefore, it becomes important to introduce some weak prior without the domain gap caused by heterogeneous data. Traditional iterative methods\cite{ van2011iterative,krumm2008reducing}  can effectively introduce prior information about the total number of materials in the object, but they cannot benefit from machine learning. Although in natural scenes, \cite{liu2023partition} train multiple INRs to learn each segmantation part and benefit from  semantic prior, the method only focus on natural image embedding and is hard to apply to CT reconstruction since accurate segmentation can’t be obtained before reconstruction. \cite{schneider2024implicit}  shows that INR can represent binary segmentation of natural images, but it requires a non-decreasing activation function and focuses solely on semantic segmentation, rather than reconstruction.

\section{Methods}
\subsection{Distribution Representation and AC Estimator}
\label{sec:Distribution Representation and AC Estimator}

For any coordinate $z \in \mathbb{R}^{n\times1}$, our method will map it to a distribution $D_z$, which characterizes the degree to which the point $p_z$  in coordinate $z$ belongs to different material:

\begin{equation}
f_{\theta}: z \mapsto D_z
\end{equation}

In industry, especially 3D printed products, the total amount of materials $K$ inside an object can be known in advance. So $D_z$ appears in the form of a discrete distribution, which is defined in the  material ordinal set $M$ with cardinal number $K$:
\begin{equation}
\begin{aligned}
D_z: M \mapsto [0,1]& \\
\text{s.t.} \sum_{m\in M} D_z(m)= 1 
\end{aligned}
\end{equation}
Where $M=\{1,2,...,K\}$ and $m \in M$. In the context of voxel discretization of CT, an intuitive understanding suggests that when $p_z$  is within a single material, the unknown ground truth of $D_z$ manifests a peak distribution. Conversely, when $p_z$ is located at the boundary between multiple materials, $D_z$ degenerates into a multi-peak form.

We then construct a function $c_{\varphi}$ controlled by a learnable parameter vector $\varphi$, to map each material oridinal $m$ in $M$ to a different linear attenuation coefficient:
\begin{equation}
c_{\varphi}: m \mapsto A_m
\end{equation}
The linear attenuation coefficient at every position in the scanned object is the objective of CT reconstruction \cite{hansen2021computed}.  Hereinafter, the linear attenuation coefficient is referred to as the attenuation coefficient (AC). However, directly establishing the mapping between position $z$ and AC through a neural network makes it hard to embed prior about the total number of material categories, thus affecting performance. The proposed distribution representation method of $f_{\theta}$ and $c_{\varphi}$ breaks down the direct mapping from $z$ to AC, allowing the information about the total number of material categories to be implicitly embedded. 

After completing the mathematical modeling of the two mappings: $f_{\theta}$ and $c_{\varphi}$, we compose the distribution $D_z$ with $c_{\varphi}(m)$ to get the AC at the location $z$:

\begin{equation}
\label{eq:intergal_Dz}
    \sum_{m\in M}D_z(m) \, c_{\varphi}(m)
\end{equation}

Therefore, the AC value located in coordinate $z$ in the CT image can be represented by  functions $g_{\varphi,\theta}$:
\begin{equation}
    g_{\theta,\varphi}(z)\overset{\triangle}{=}\sum_{m\in M}[f_{\theta}(z)](m) \, c_{\varphi}(m) 
\label{eq:also-important}
\end{equation}

To ensure $g_{\theta, \varphi}(\cdot)$ is differentiable and  capable of representing the high-frequency information in the CT image, we will construct appropriate functions $f_{\theta}(\cdot)$ and $c_{\varphi}(\cdot)$ as below. For $f_{\theta}(\cdot)$, inspired by ~\cite{sitzmann2020implicit} and ~\cite{tancik2020fourfeat}, firstly, we present a random matrix $E\in \mathbb{R}^{m\times n}$ with trigonometric functions to map the coordinate $z\in \mathbb{R}^{n\times 1}$ to a vector $r \in \mathbb{R}^{2m\times 1}$ representing a series random Fourier basis:

\begin{equation}
r \stackrel{\triangle}{=}
\begin{pmatrix}
    \sin(E \times z) \\
    \cos(E \times z)
\end{pmatrix}
\end{equation}

The elements in the random matrix $E$ are independently sampled from $\mathcal{N}(0,\sigma^2)$ and frozen during algorithm training similar to \cite{tancik2020fourfeat}.
For the point $r$ on the hypersphere 
$S^{2m-1}$ formed by Fourier basis, we construct an MLP with $L$ layers, where the $1^{st}$ to $(L-1)^{th}$ layers use the sinusoidal function as the activation function:
\begin{equation}
    u_{i+1}=\sin(W_iu_i+b_i),
    i=1,2,...,L-1
\end{equation}
Among them, the input $u_1$ of the first layer of the neural network is Fourier basis vector $r$. Here, $W_i \in \mathbb{R}^{l_{i+1}\times l_i}$, $b_i \in \mathbb{R}^{l_{i+1}\times 1}$, with $l_1=2m$ and $l_L=|M|$.

To ensure that $f_{\theta}$ can output a probability distribution of $|M|$ classes and tends to have a steep unimodal distribution, we use an modulated Softmax function control by $T$ to construct the $L^{th}$ layer, and only use  $|M|$ neurons:

\begin{equation}
\begin{aligned}
    u_{L+1}&=\text{Sofmax}(W_Lu_L+b_L), T)\\
\end{aligned}
\end{equation}

More specifically, the expanded formula of above is:
\begin{equation}
\begin{aligned}
    u_{L+1}[i,:]&= \frac{e^{z_i / T}}{\sum_{j=1}^{l_L} e^{z_j / T}}\\
\end{aligned}
\end{equation}

\begin{figure}[ht]
    \centering
    \includegraphics[width=8cm]{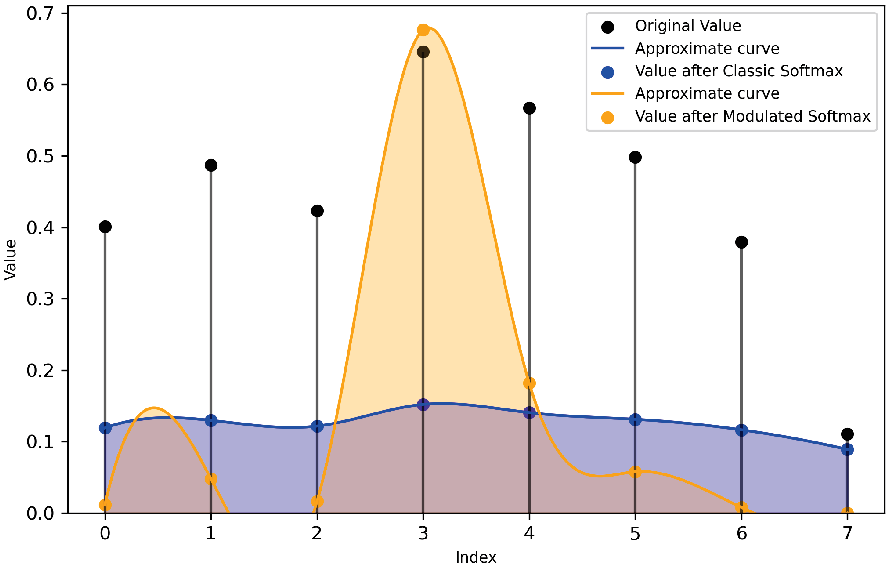} 
    \caption{Comparison between modulated softmax and classic softmax, where the modulation coefficient $T$ is equal to 0.06.}
    \label{fig:modulated_softmax}
\end{figure}

Among them, $z_i = (W_Lu_L+b_L)[i]$, $i=1,2,...,l_L$. Modulated Softmax with a coefficient $T\in(0,1)$ will make the distribution steeper compared to classical Softmax, as shown in \cref{fig:modulated_softmax}. Thereby, it's more closely approximating the internal structure of the object, where most of the coordinates are located within a single material area, rather than at the boundary of multiple materials. As \cref{sec:Dynamics Analysis During Optimization} shows, by using the modulation way, our method can learn the segmentation effectively in an unsupervised manner.

 For $c_{\varphi}(\cdot)$, although it can be chosen in a very large function space $\mathcal{F}\stackrel{\triangle}{=}\{\tilde{c}_{\varphi}||\varphi|\in \mathbb{N}_0\}$, we restrict the function space $\mathcal{F}$ to a subspace $\mathcal{F}_{|M|}\stackrel{\triangle}{=}\{\tilde{c}_{\varphi}||\varphi|= |M|\}$,  this work specifically selects $c_{\varphi}\in \mathcal{F}_{|M|}$ with the specific compact analytical form:
\begin{equation}
c_{\varphi}: m_{i} \mapsto \varphi_{i}
\end{equation}
Among them, $i=1,2,...,|M|$ and $m_{i}\in M$. According to its role, we name $c_{\varphi}$ as the AC estimator. If more material information were available, choosing a function in $\mathcal{F}$ with an output range within a specific interval could improve algorithm performance. However, it's not within the scope of this work because we assume that we can't obtain additional information except the total number of material categories.

After constructing $f_{\theta}(\cdot)$ and $c_{\varphi}(\cdot)$,  the $g_{\theta,\varphi}(\cdot)$ can be obtained from \cref{eq:also-important} accordingly.  We will show that this compact functional form will have good performance for sparse CT reconstruction tasks in \cref{sec:Performance Comparison}. And it shows dynamic characteristics with clear physical meaning in \cref{sec:Dynamics Analysis During Optimization}. Next we will  design two methods in a progressive manner to initialize the parameters $\theta,\varphi$ in \cref{sec:Initialization Strategy}.

\subsection{Initialization Strategy}
\label{sec:Initialization Strategy}
Before jointly optimizing $\theta$ in $f_{\theta}$ and $\varphi$ in $c_{\varphi}$, we design two different methods to estimate their initial value.

In the first method, for initialization of $\theta$, we use a method introduced in \cite{sitzmann2020implicit}. For initialization of $\varphi$, we use the rapid reconstruction method FBP to perform an initial reconstruction, and then use a segmentation method to segment different material areas, and finally calculate the mean attenuation coefficients of different segmented areas.

Specifically, We used the method $f_t$ ($f_t$=FBP in our AC-IND) to reconstruct the object from sinogram $y$, then use Multi-Otsu Thresholding method to generate masks to divide the image into $|M|$ non-intersecting regions: 
\begin{figure*}[t]
    \centering
    \includegraphics[width=\textwidth]{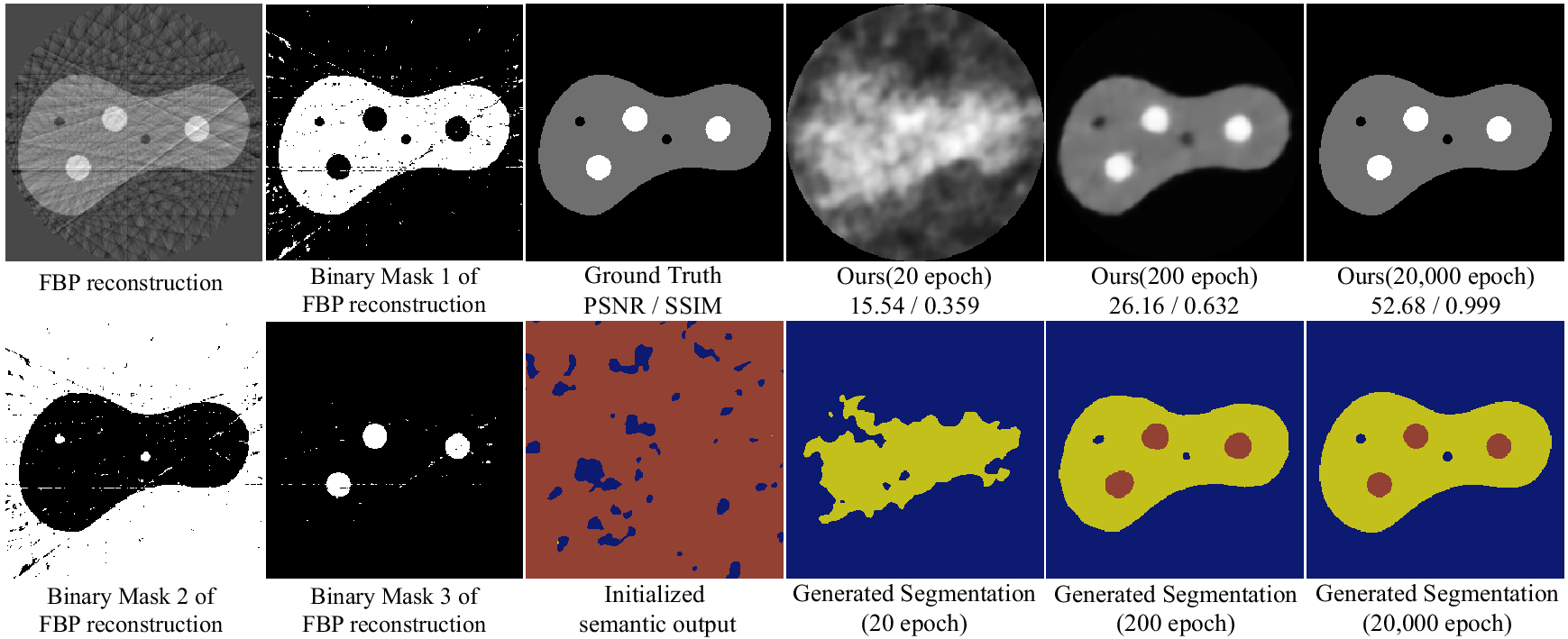}
    \caption{The first row: FBP reconstrcution, Mask, the Ground Truth of Barbapapa phantom and the reconstruction from 20 views by our method. The second row: Mask, segmentation output automatically produced by our method during the reconstruction process.}
    \label{fig:FBP_mask_barbapapa_recon_seg}
\end{figure*}

\begin{equation}
\begin{aligned}
\text{Mask}_j=f_{\text{ostu}}(f_t(y),|M|)_j
\end{aligned}
\end{equation}
where $\text{Mask}_j$($j=1,2,...,|M|$) is a binary matrix, a visulization exmaple of which is shown in \cref{fig:FBP_mask_barbapapa_recon_seg}. 

Finally, we calculate the mean $a_j$ of the rough segmentation regions, which is used as the initialization of the differentiable parameter $\varphi_j$ of the function $c_{\varphi}$:
\begin{equation}
\begin{aligned}
a_j\stackrel{\triangle}{=}\mu(f_t(y) \odot \text{Mask}_j)
\end{aligned}
\end{equation}
where $\odot$ represents the Hadamard product and $\mu(\cdot)$  represents a averaging operation for non-zero elements. We named our algorithm based on the initialization strategy as AC-IND.

Intuitively, the more precise the initial estimated value of $\varphi$ is, the more precise AC will be learned during the joint optimization of $f_{\theta}$ and $c_{\varphi}$. In view of this, we developed a second initialization method. For initializing $\theta$, we use the same method as the AC-IND. For initializing $\varphi$, we first apply AC-IND to obtain a reconstruction, then segment it using the Multi-Otsu Thresholding method. Finally, we calculate the mean values of the segmented regions and use these means as the initial values of $\varphi_j$. Essentially, the second method replaces $f_t$ from FBP with AC-IND. We named the algorithm based on the enhanced initialization strategy as AC-IND\textsuperscript{+}. \cref{sec:Performance Comparison} will present the performance of the AC-IND and highlight the improvement of AC-IND\textsuperscript{+}.

\subsection{Loss Function}
The loss function is defined based on the $L_2$ norm:
\label{sec:Loss Function}
\begin{equation}
\begin{aligned}
L_{\theta,\varphi}&=\|AX_{\theta,\varphi}-y\|_2
\end{aligned}
\end{equation}

$A: \mathbb{R}^{H\times W}\rightarrow \mathbb{R}^{U\times V}$  is an operator equivalent to the discrete representation of the Radon transform. It is determined by the internal parameters of the machine during CT scanning. $X_{\theta,\varphi}\in \mathbb{R}^{H\times W}$ is the object to be reconstructed:  
\begin{equation}
\begin{aligned}
X_{\theta,\varphi}[i,j]=\sum_{k=1}^{|M|}f_{\theta}((i,j))(k)c_{\varphi}(k)\\
i=1,2,...,H, j=1,2,...,W
\end{aligned}
\end{equation}
$y\in \mathbb{R}^{U\times V}$ is the collected sinogram (measurements) which indicates that the
CT machine acquires $U$ projections at different angles and the detector has $V$ detector pixels. 

\section{Experiments}
We conducted a dynamics analysis of AC-IND's optimization process. In addition, we also conducted quantitative and qualitative comparisons with other methods.
\subsection{Dataset and pre-processing}
For dynamics analysis in \cref{sec:Dynamics Analysis During Optimization}, we construct a $256 \times 256$ phantom containing three materials based on the Barbapapa phantom \cite{van2011iterative, schoonhoven2024auto}.  The three materials are polymethyl methacrylate, aluminium, air. A benefit of having only three materials is that it allows us to fully visualize the learned 3D AC vector in a 3D Cartesian coordinate system. We generate projection data from 20 views and utilize our AC-IND to reconstruct the CT image from these projections. 

For Performance Comparison in \cref{sec:Performance Comparison}, we create a Walnut Slice Dataset extracted from 3D Walnut scans, and create an Ellipse Material Dataset using a Shepp-Logan\cite{shepp1974fourier} generator. We first collect 3D volumes numbered 1 to 8 from the 3D Walnut Dataset \cite{der2019cone}. Then, for each individual walnut volume, we generate 10 different slices, so a total of 80 different slices is produced. For each slice, we generate sparse projection data at 20, 40 and 60 views.

We use a Sheep-Logan generator\footnote{https://pypi.org/project/phantominator/} to form a combinations of industrial items containing 5 materials and an environmental material consisting of air. Finally, we created a dataset containing 80 samples and named it the Ellipse Material Dataset. Similar to the Walnut Slice Dataset, we generated sparse projection data at 20, 40 and 60 views.

\subsection{Implementation Details}

The linear operator $A$ is implemented by Tomosipo \cite{hendriksen2021tomosipo}. The geometry of the CT setup is parallel beam. During scanning, the radiation source will rotate equiangularly around the center of the object. For all datasets, we used the Adam algorithm with $(\beta_1, \beta_2)$ equal to $(0.9, 0.999)$ to optimize the trainable parameters. For the Walnut Slice Dataset, we set the learning rates of all layers of the MLP to $4.0\times10^{-5}$ and set the learning rates of the AC estimator to $1.0\times10^{-5}$. For the  Ellipse Material Dataset, we set the learning rates of all layers of the MLP and parameters of the AC estimator to $1.0\times10^{-4}$. The modulation coefficients of Softmax for the Walnut Slice Dataset and the Ellipse Material Dataset are set to 0.2 and 0.035, respectively. For both datasets, the variance of the zero-mean Gaussian distribution used to construct the random matrix $E$ is 16.0.

\subsection{Dynamics Analysis During Optimization}
\label{sec:Dynamics Analysis During Optimization}
\begin{figure}[ht]
    \centering
    \includegraphics[width=8cm]{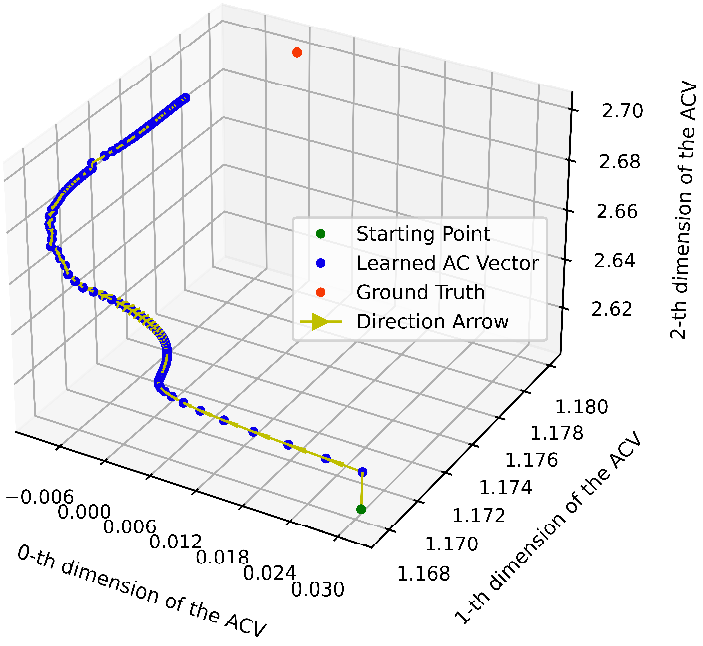} 
    \caption{Learned Attenuation Coefficient Vector  (ACV)'s trajectory during training. The initial ACV is in the lower right corner.}
    \label{fig:3D-plot-of-leaned-acv}
\end{figure}

\begin{figure}[t]
    \centering
    \includegraphics[width=8cm]{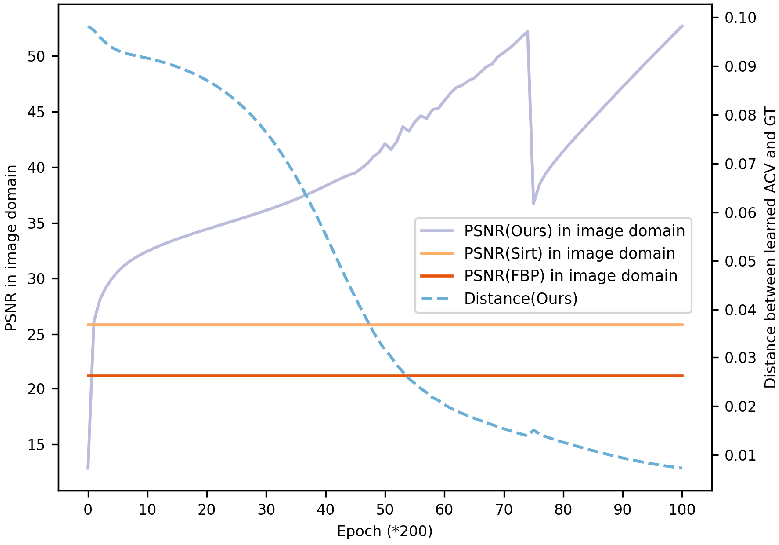} 
    \caption{Reconstruction accuracy of our method and compared methods (solid line, based on left Y-axis); Distance between attenuation coefficient vector learned by our method and the ground truth (\textcolor{lightblue}{blue} dotted line, based on right Y-axis).}
    \label{fig:epoch-psnr-acv}
\end{figure}
In this section, we will illustrate that our method's intermediate output exhibits clear physical meaning:

\textbf{I. Our method automatically generates segmentation maps during reconstruction}.  Specifically, during the neural network training process, we calculate the pixel-level classification $\text{Label}_z$ for every coordinate $z$ as below:  
\begin{equation}
\begin{aligned}
\text{Label}_z=\underset {m\in M}{\text{Argmax}}(f_{\theta}(z)(m))
\end{aligned}
\end{equation}
By traversing the coordinate $z \in \{0/H,...,(H-1)/H\} \times \{0/W,...,(W-1)/W\}$,  we can get a segmentation map $S\in \{1,2,...,|M|\}^{H\times W}$.

As shown in the \cref{fig:FBP_mask_barbapapa_recon_seg}, as the reconstruction accuracy improves, the segmentation map $S$ represented by the intermediate output of our algorithm also becomes more accurate. This phenomenon indicates that our method can perform unsupervised segmentation in the image domain concurrently with CT reconstruction.

\textbf{II. The learned AC Vector (ACV) of the AC estimator gradually approaches the ACV’s ground truth}. As \cref{fig:epoch-psnr-acv} shows, the AC estimator $c_{\varphi}$ approximates the true ACV using only the projection data's supervision, without relying on any information from the ACV's ground truth. Additionally, \cref{fig:3D-plot-of-leaned-acv} depicts the movement of the learned ACV in three-dimensional space. It can be observed that it gradually converges toward the ACV's ground truth. The pattern of the movement suggests that the learned ACV first approaches the real ACV in one dimension, and then approaches in other dimensions in turn.
\subsection{Performance Comparison}
\label{sec:Performance Comparison}
\begin{table}[hb]
\caption{Quantitative evaluation on Walnut Slice Dataset and expressed as mean $\pm$ standard  deviation. The best value is \textbf{bolded}. }
\label{tab:Walnut numerical}
\small
\centering
\begin{tabular}{p{0.4cm} c| c c c}
\toprule
 & & 20 views & 40 views & 60 views \\
\hline
\multirow{2}{*}{FBP} & PSNR & 15.31$\pm$1.00 &  21.24$\pm$1.04 & 25.03$\pm$1.04 \\

                     & SSIM & 0.344$\pm$0.010 &  0.449$\pm$0.018 & 0.554$\pm$0.025 \\
\hline

\multirow{2}{*}{SIRT} & PSNR & 19.78$\pm$0.95 & 23.76$\pm$1.03 & 26.45$\pm$1.05 \\
                      & SSIM & 0.393$\pm$0.020 &  0.545$\pm$0.029 & 0.681$\pm$0.030 \\
\hline
\multirow{2}{*}{INR} & PSNR & 23.80$\pm$1.32 &  30.72$\pm$1.13 & 34.87$\pm$0.93 \\
                     & SSIM & 0.541$\pm$0.036 &  0.789$\pm$0.024 & 0.909$\pm$0.012 \\
\hline
\multirow{2}{*}{Ours} & PSNR & \cellcolor{gray!20} \textbf{26.49}$\pm$2.43 & \cellcolor{gray!20}  \textbf{35.38}$\pm$1.05 & \cellcolor{gray!20} \textbf{37.66}$\pm$0.89 \\
                      & SSIM & \cellcolor{gray!20} \textbf{0.752}$\pm$0.120 & \cellcolor{gray!20}  \textbf{0.955}$\pm$0.008 & \cellcolor{gray!20} \textbf{0.967}$\pm$0.007 \\
\bottomrule
\end{tabular}
\end{table}

\begin{figure*}[t]
    \centering
    \includegraphics[width=\textwidth]{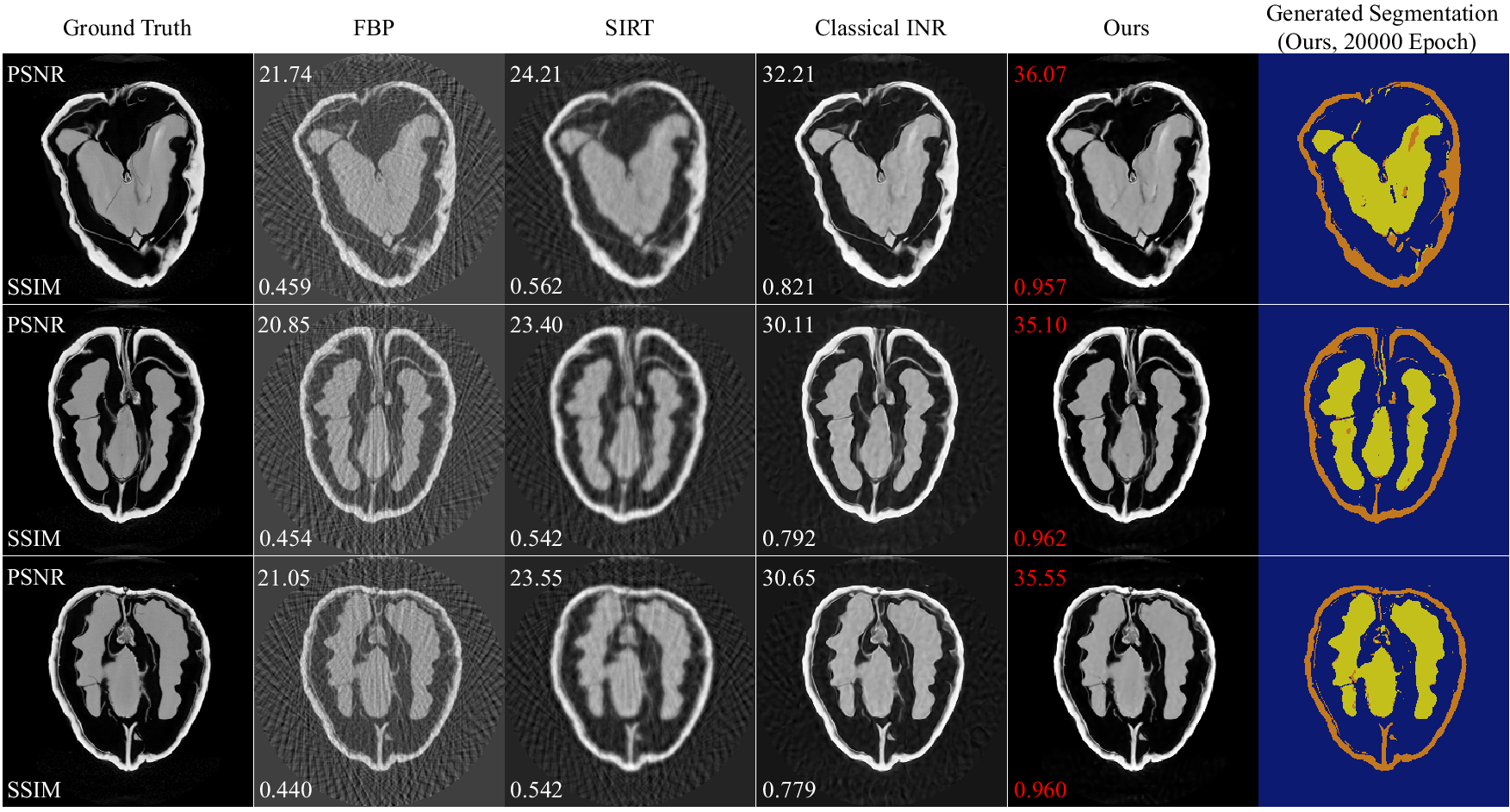} 
    \caption{Different reconstruction results under 40 views setting(zoom in to see details), and the segmentation generated by our method.}
    \label{fig:walnut_recon}
\end{figure*}
\begin{figure*}[ht]
    \centering
    \includegraphics[width=\textwidth]{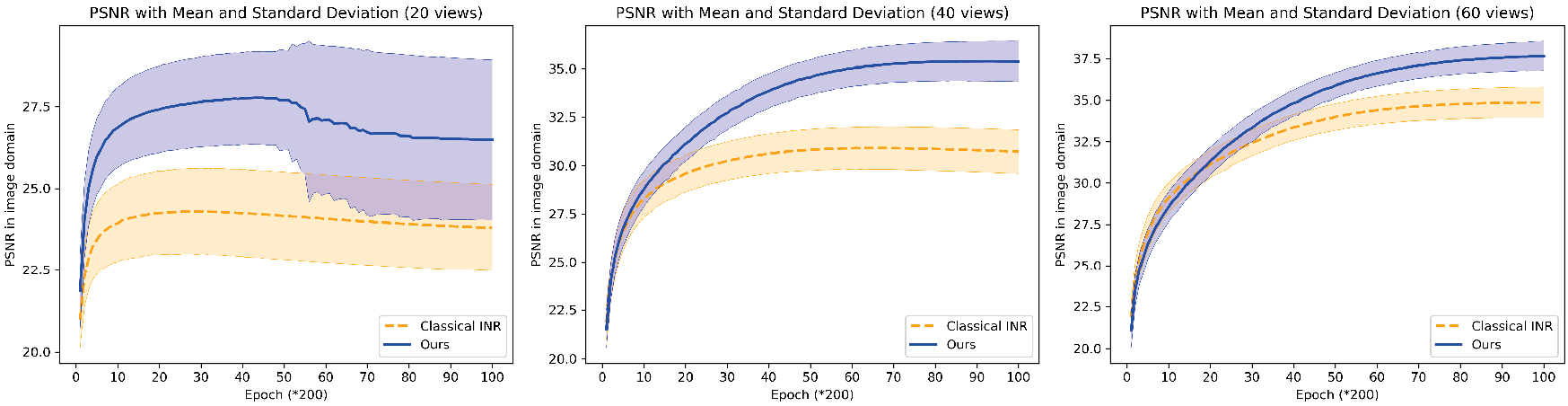} 
    \caption{The average PSNR and the corresponding standard deviation of the entire Wulnut Slice Dataset in each epoch.}
    \label{fig:walnut_epoch_psnr}
\end{figure*}

\begin{figure*}[t]
    \centering
    \includegraphics[width=\textwidth]{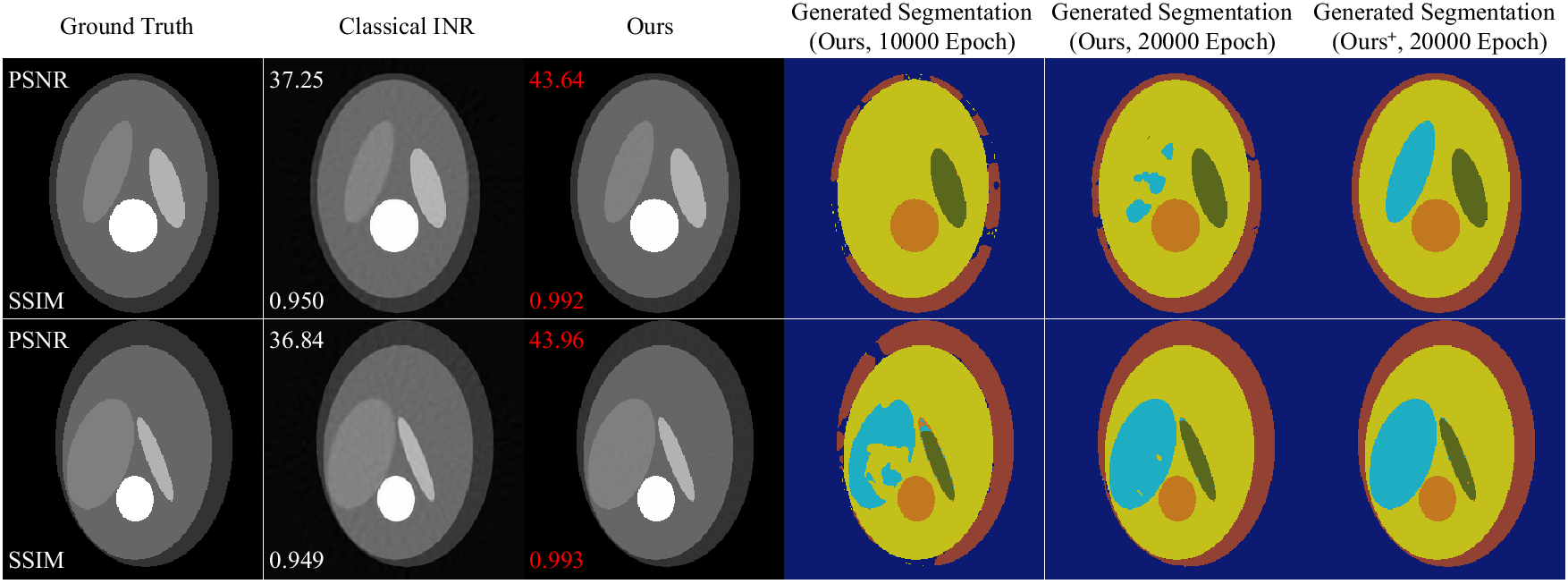} 
    \caption{Different reconstruction results under 40 views setting(zoom in to see details), and the segmentation generated by ours and ours\textsuperscript{+}.}
    \label{fig:ellipse_recon}
\end{figure*}

We conducted experiments on the Walnut Slice Dataset and Ellipse Material Dataset with a total number of sparse views of 20, 40, and 60. Although a walnut is mainly composed of three substances: air, shell, and pulp, it is not a man-made industrial product and is composed of more complex materials. In this challenging dataset, we empirically assume it contains 6 different types of materials. We will compare our AC-IND with other methods: classic INR using same random Fourier feature as AC-IND, FBP and SIRT. The number of iterations of SIRT is set as 30,000.

\textbf{I. Walnut Slice Dataset}, \cref{tab:Walnut numerical} shows that in the case of sparse reconstruction at 20 views, 40 views and 60 views, our method outperformed classical INR by 2.69 dB, 4.66 dB and 2.79 dB, respectively.  \cref{fig:walnut_recon} shows our reconstruction results have fewer artifacts and sharper edges compared to classical INR. Additionally, \cref{fig:walnut_recon} also displays the semantic segmentation map that is automatically generated by our method. Although there is no ground truth of the segmentation so that the Jaccard index and Dice index cannot be calculated, it shows a good visual segmentation result.  The above results demonstrate that our method not only has better reconstruction performance than the competitive method, but also can complete the unsupervised segmentation at the same time(see also the Appendix). For both the classical INR and our AC-IND, we report the results at the 20,000th epoch in \cref{tab:Walnut numerical}. Additionally, for each epoch, we visualized the PSNR value of the two methods in \cref{fig:walnut_epoch_psnr} and found that \textbf{our method consistently outperformed the classic INR in nearly every epoch}.

\textbf{II. Ellipse Material Dataset}, as shown in \cref{tab:Ellipse Material numerical}, in the case of sparse reconstruction at 20 views, 40 views and 60 views, our method exceeded classical INR by 2.69 dB, 5.4 dB and 6.58 dB respectively. For fair comparison, the numerical results of classical INR at epoch 3,400 is reported, when classical INR usually reaches its maximum peak and after then its performance starts to degrade. The numerical performance of our work at epoch 10,000 is reported. \cref{fig:ellipse_recon} shows our method's reconstruction visualization at 10,000 epoch under 40 views and the automatically generated semantic segmentation by our method at 10,000 epoch and 20,000 epochs. It shows that our method has fewer artifacts than the contrast method and the segmentation is improved as the training epoch increases (see also the Appendix).

\textbf{III. Optimizing initialization of AC estimator.} Because the Ellipse Material Dataset contains ellipses with similar attenuation coefficients, the semantic segmentation task is more difficult for our AC-IND. To improve it, we change the $f_t$ in \cref{sec:Initialization Strategy} from FBP to AC-IND to form our new method: AC-IND\textsuperscript{+} (Ours\textsuperscript{+}). The substitution of $f_t$ naturally provides the AC estimator with a better initial value, as AC-IND is much better than FBP in the reconstruction task.
\cref{tab:Ellipse Material Dataset newcompare} shows that AC-IND\textsuperscript{+} outperforms AC-IND by 5.4 dB in the case of 20 views reconstruction. And AC-IND\textsuperscript{+} exceeds AC-IND by 3.26 dB and 2.35 dB in the case of 40 view and 60 view reconstruction, respectively. \cref{fig:ellipse_recon} shows that AC-IND\textsuperscript{+} can generate better segmentation images than AC-IND in the same number of training epochs. Logically speaking, AC-IND\textsuperscript{+} is equivalent to the iterative extension of AC-IND based on the change of $f_t$. The better performance of AC-IND\textsuperscript{+} compared to AC-IND highlights the critical importance of the AC estimation in sparse CT reconstruction and unsupervised segmentation. 
\begin{table}[t]
\caption{Quantitative evaluation on Ellipse Material Dataset and expressed as mean $\pm$ standard  deviation. The best value is \textbf{bolded}.}
\label{tab:Ellipse Material numerical}
\small
\centering
\begin{tabular}{p{0.4cm} c| c c c}
\toprule
 & & 20 views & 40 views & 60 views \\
\hline
\multirow{2}{*}{FBP} & PSNR & 23.33$\pm$0.40 &  28.51$\pm$0.38 & 31.62$\pm$0.37 \\

                     & SSIM & 0.386$\pm$0.012 &   0.556$\pm$0.016 &  0.707$\pm$0.015 \\
\hline

\multirow{2}{*}{SIRT} & PSNR & 28.10$\pm$0.35 &  31.14$\pm$0.35 & 32.99$\pm$0.35 \\
                      & SSIM & 0.690$\pm$0.019 &  0.804$\pm$0.013 & 0.869$\pm$0.009 \\
\hline
\multirow{2}{*}{INR} & PSNR & 31.95$\pm$2.48 &  36.57$\pm$3.17 & 37.67$\pm$2.09 \\
                     & SSIM & 0.812$\pm$0.061 &  0.931$\pm$0.109 & 0.965$\pm$0.074 \\
\hline
\multirow{2}{*}{Ours} & PSNR & \cellcolor{gray!20} \textbf{34.64}$\pm$6.67 &  \cellcolor{gray!20} \textbf{41.97}$\pm$3.76 & \cellcolor{gray!20} \textbf{44.25}$\pm$1.11 \\
                      & SSIM & \cellcolor{gray!20} \textbf{0.882}$\pm$0.171 & \cellcolor{gray!20}  \textbf{0.972}$\pm$0.048 & \cellcolor{gray!20} \textbf{0.993}$\pm$0.004 \\
\bottomrule
\end{tabular}
\end{table}

\begin{table}[t]
\caption{Quantitative comparison of AC-IND and AC-IND\textsuperscript{+} on the Ellipse Material Dataset, expressed as mean ± standard deviation.}
\label{tab:Ellipse Material Dataset newcompare}
\small
\centering
\begin{tabular}{p{0.4cm} c| c c c}
\toprule
 & & 20 views & 40 views & 60 views \\
 \hline
{Ours} & PSNR &  34.64$\pm$6.67 &   41.97$\pm$3.76 &  44.25$\pm$1.11 \\
                      \scalebox{0.6}{(AC-IND)}& SSIM & 0.882$\pm$0.171 &   \textbf{0.972}$\pm$0.048 & \textbf{0.993}$\pm$0.004 \\
\hline
{Ours\textsuperscript{+}} & PSNR & \cellcolor{gray!20} \textbf{40.04}$\pm$9.39 &  \cellcolor{gray!20} \textbf{45.23}$\pm$5.37 & \cellcolor{gray!20} \textbf{46.60}$\pm$5.02 \\
                      \scalebox{0.6}{(AC-IND\textsuperscript{+})} & SSIM & \cellcolor{gray!20} \textbf{0.902}$\pm$0.201 & \cellcolor{gray!20}  \textbf{0.972}$\pm$0.088 & \cellcolor{gray!20} 0.982$\pm$0.088 \\

\bottomrule
\end{tabular}
\end{table}

\begin{table}[htb]
    \centering
    \caption{Comparison of the trainable parameters and FLOPs}
    \label{tab:Comparison of Parameters and FLOPs}
    \begin{tabular}{>{\centering\arraybackslash}p{2.2cm} | >{\centering\arraybackslash}p{2.2cm} >{\centering\arraybackslash}p{2.2cm}}
        \toprule
        \textbf{Method} & \textbf{Parameters$\downarrow$} & \textbf{FLOPs$\downarrow$} \\
        \midrule
        INR  & 460.55 k & 30.18 GMac \\
        Ours & \cellcolor{gray!20}\textbf{396.04} k & \cellcolor{gray!20}\textbf{26.06} GMac  \\      
        \bottomrule
     \end{tabular}
\end{table}

\textbf{IIII. Comparison of model parameters}. As \cref{tab:Comparison of Parameters and FLOPs} shows, the number of  trainable parameters and Floating Point Operations(FLOPs) of our algorithm is less than that of the classic INR, which means that our method requires fewer parameters while achieving better performance. 

\section{Conclusion}
We propose AC-IND to compute sparse CT reconstruction and automatically generate segmentations. Unlike traditional INR, which maps coordinates to a scalar, our algorithm maps them to a distribution. Furthermore, we design an AC estimator and initialize it with the mean of the segmented region, which is calculated from a rough reconstructed image by a fast reconstruction method. In joint optimization of the distribution and the AC estimator, as the reconstruction accuracy improves, the generated segmentation becomes more accurate, and the AC estimator's output converges to the AC's ground truth. Experiments demonstrate that AC-IND outperforms the comparative method in the reconstruction task. Additionally, the enhanced version AC-IND\textsuperscript{+} of AC-IND demonstrates that a better initialization of the AC estimator facilitates reconstruction. Importantly, our method shows reconstruction and segmentation can be completed simultaneously. \\
\textbf{Acknowledgements}. This research work was undertaken in the context of the
Horizon 2020 MSCA ITN project “xCTing” (Project ID: 956172). We acknowledge funding from the Flemish Government (AI Research Program) and the Research Foundation - Flanders (FWO) through project number G0G2921N. Richard Schoonhoven was funded by the Dutch Research Council (NWO) in the framework of the NWA-ORC Call (file number NWA.1160.18.316). 

{\small
\bibliographystyle{ieee_fullname}
\bibliography{egbib}

\begin{thebibliography}{10}\itemsep=-1pt

\bibitem{andersen1984simultaneous}
Anders~H Andersen and Avinash~C Kak.
\newblock Simultaneous algebraic reconstruction technique (sart): a superior implementation of the art algorithm.
\newblock {\em Ultrasonic imaging}, 6(1):81--94, 1984.

\bibitem{cai2024batch}
Zhicheng Cai, Hao Zhu, Qiu Shen, Xinran Wang, and Xun Cao.
\newblock Batch normalization alleviates the spectral bias in coordinate networks.
\newblock In {\em Proceedings of the IEEE/CVF Conference on Computer Vision and Pattern Recognition}, pages 25160--25171, 2024.

\bibitem{chambolle2004algorithm}
Antonin Chambolle.
\newblock An algorithm for total variation minimization and applications.
\newblock {\em Journal of Mathematical imaging and vision}, 20:89--97, 2004.

\bibitem{der2019cone}
Henri Der~Sarkissian, Felix Lucka, Maureen van Eijnatten, Giulia Colacicco, Sophia~Bethany Coban, and Kees~Joost Batenburg.
\newblock A cone-beam x-ray computed tomography data collection designed for machine learning.
\newblock {\em Scientific data}, 6(1):215, 2019.

\bibitem{fessler1994penalized}
Jeffrey~A Fessler.
\newblock Penalized weighted least-squares image reconstruction for positron emission tomography.
\newblock {\em IEEE transactions on medical imaging}, 13(2):290--300, 1994.

\bibitem{gordon1970algebraic}
Richard Gordon, Robert Bender, and Gabor~T Herman.
\newblock Algebraic reconstruction techniques (art) for three-dimensional electron microscopy and x-ray photography.
\newblock {\em Journal of theoretical Biology}, 29(3):471--481, 1970.

\bibitem{gu2023generalizable}
Jeffrey Gu, Kuan-Chieh Wang, and Serena Yeung.
\newblock Generalizable neural fields as partially observed neural processes.
\newblock In {\em Proceedings of the IEEE/CVF International Conference on Computer Vision}, pages 5330--5339, 2023.

\bibitem{han2016deep}
Yo~Seob Han, Jaejun Yoo, and Jong~Chul Ye.
\newblock Deep residual learning for compressed sensing ct reconstruction via persistent homology analysis.
\newblock {\em arXiv preprint arXiv:1611.06391}, 2016.

\bibitem{hansen2021computed}
Per~Christian Hansen, Jakob J{\o}rgensen, and William~RB Lionheart.
\newblock {\em Computed tomography: algorithms, insight, and just enough theory}.
\newblock SIAM, 2021.

\bibitem{he2020radon}
Ji He, Yongbo Wang, and Jianhua Ma.
\newblock Radon inversion via deep learning.
\newblock {\em IEEE transactions on medical imaging}, 39(6):2076--2087, 2020.

\bibitem{hendriksen2021tomosipo}
Allard~A Hendriksen, Dirk Schut, Willem~Jan Palenstijn, Nicola Vigan{\'o}, Jisoo Kim, Dani{\"e}l~M Pelt, Tristan Van~Leeuwen, and K~Joost Batenburg.
\newblock Tomosipo: fast, flexible, and convenient 3d tomography for complex scanning geometries in python.
\newblock {\em Optics Express}, 29(24):40494--40513, 2021.

\bibitem{herman2009fundamentals}
Gabor~T Herman.
\newblock {\em Fundamentals of computerized tomography: image reconstruction from projections}.
\newblock Springer Science \& Business Media, 2009.

\bibitem{jacot2018neural}
Arthur Jacot, Franck Gabriel, and Cl{\'e}ment Hongler.
\newblock Neural tangent kernel: Convergence and generalization in neural networks.
\newblock {\em Advances in neural information processing systems}, 31, 2018.

\bibitem{jin2017deep}
Kyong~Hwan Jin, Michael~T McCann, Emmanuel Froustey, and Michael Unser.
\newblock Deep convolutional neural network for inverse problems in imaging.
\newblock {\em IEEE transactions on image processing}, 26(9):4509--4522, 2017.

\bibitem{kazerouni2024incode}
Amirhossein Kazerouni, Reza Azad, Alireza Hosseini, Dorit Merhof, and Ulas Bagci.
\newblock Incode: Implicit neural conditioning with prior knowledge embeddings.
\newblock In {\em Proceedings of the IEEE/CVF Winter Conference on Applications of Computer Vision}, pages 1298--1307, 2024.

\bibitem{krumm2008reducing}
Michael Krumm, Stefan Kasperl, and Matthias Franz.
\newblock Reducing non-linear artifacts of multi-material objects in industrial 3d computed tomography.
\newblock {\em Ndt \& E International}, 41(4):242--251, 2008.

\bibitem{liao2001fast}
Ping-Sung Liao, Tse-Sheng Chen, Pau-Choo Chung, et~al.
\newblock A fast algorithm for multilevel thresholding.
\newblock {\em J. Inf. Sci. Eng.}, 17(5):713--727, 2001.

\bibitem{lin2023learning}
Yiqun Lin, Zhongjin Luo, Wei Zhao, and Xiaomeng Li.
\newblock Learning deep intensity field for extremely sparse-view cbct reconstruction.
\newblock In {\em International Conference on Medical Image Computing and Computer-Assisted Intervention}, pages 13--23. Springer, 2023.

\bibitem{liu2023partition}
Ke Liu, Feng Liu, Haishuai Wang, Ning Ma, Jiajun Bu, and Bo Han.
\newblock Partition speeds up learning implicit neural representations based on exponential-increase hypothesis.
\newblock In {\em Proceedings of the IEEE/CVF International Conference on Computer Vision}, pages 5474--5483, 2023.

\bibitem{mildenhall2021nerf}
Ben Mildenhall, Pratul~P Srinivasan, Matthew Tancik, Jonathan~T Barron, Ravi Ramamoorthi, and Ren Ng.
\newblock Nerf: Representing scenes as neural radiance fields for view synthesis.
\newblock {\em Communications of the ACM}, 65(1):99--106, 2021.

\bibitem{rahaman2019spectral}
Nasim Rahaman, Aristide Baratin, Devansh Arpit, Felix Draxler, Min Lin, Fred Hamprecht, Yoshua Bengio, and Aaron Courville.
\newblock On the spectral bias of neural networks.
\newblock In {\em International conference on machine learning}, pages 5301--5310. PMLR, 2019.

\bibitem{rahimi2007random}
Ali Rahimi and Benjamin Recht.
\newblock Random features for large-scale kernel machines.
\newblock {\em Advances in neural information processing systems}, 20, 2007.

\bibitem{rematas2022urban}
Konstantinos Rematas, Andrew Liu, Pratul~P Srinivasan, Jonathan~T Barron, Andrea Tagliasacchi, Thomas Funkhouser, and Vittorio Ferrari.
\newblock Urban radiance fields.
\newblock In {\em Proceedings of the IEEE/CVF Conference on Computer Vision and Pattern Recognition}, pages 12932--12942, 2022.

\bibitem{ruckert2022neat}
Darius R{\"u}ckert, Yuanhao Wang, Rui Li, Ramzi Idoughi, and Wolfgang Heidrich.
\newblock Neat: Neural adaptive tomography.
\newblock {\em ACM Transactions on Graphics (TOG)}, 41(4):1--13, 2022.

\bibitem{saragadam2023wire}
Vishwanath Saragadam, Daniel LeJeune, Jasper Tan, Guha Balakrishnan, Ashok Veeraraghavan, and Richard~G Baraniuk.
\newblock Wire: Wavelet implicit neural representations.
\newblock In {\em Proceedings of the IEEE/CVF Conference on Computer Vision and Pattern Recognition}, pages 18507--18516, 2023.

\bibitem{sauer1993local}
Ken Sauer and Charles Bouman.
\newblock A local update strategy for iterative reconstruction from projections.
\newblock {\em IEEE Transactions on Signal Processing}, 41(2):534--548, 1993.

\bibitem{schneider2024implicit}
Jan~Philipp Schneider, Mishal Fatima, Jovita Lukasik, Andreas Kolb, Margret Keuper, and Michael Moeller.
\newblock Implicit representations for constrained image segmentation.
\newblock In {\em Forty-first International Conference on Machine Learning}, 2024.

\bibitem{schoonhoven2024auto}
Richard Schoonhoven, Alexander Skorikov, Willem~Jan Palenstijn, Dani{\"e}l~M Pelt, Allard~A Hendriksen, and K~Joost Batenburg.
\newblock How auto-differentiation can improve ct workflows: classical algorithms in a modern framework.
\newblock {\em Optics Express}, 32(6):9019--9041, 2024.

\bibitem{shen2022nerp}
Liyue Shen, John Pauly, and Lei Xing.
\newblock Nerp: implicit neural representation learning with prior embedding for sparsely sampled image reconstruction.
\newblock {\em IEEE Transactions on Neural Networks and Learning Systems}, 35(1):770--782, 2022.

\bibitem{shepp1974fourier}
Lawrence~A Shepp and Benjamin~F Logan.
\newblock The fourier reconstruction of a head section.
\newblock {\em IEEE Transactions on nuclear science}, 21(3):21--43, 1974.

\bibitem{shi2024implicit}
Jiayang Shi, Junyi Zhu, Daniel~M. Pelt, K.~Joost Batenburg, and Matthew~B. Blaschko.
\newblock Implicit neural representations for robust joint sparse-view {CT} reconstruction.
\newblock {\em Transactions on Machine Learning Research}, 2024.

\bibitem{sitzmann2020implicit}
Vincent Sitzmann, Julien Martel, Alexander Bergman, David Lindell, and Gordon Wetzstein.
\newblock Implicit neural representations with periodic activation functions.
\newblock {\em Advances in neural information processing systems}, 33:7462--7473, 2020.

\bibitem{song2023piner}
Bowen Song, Liyue Shen, and Lei Xing.
\newblock Piner: Prior-informed implicit neural representation learning for test-time adaptation in sparse-view ct reconstruction.
\newblock In {\em Proceedings of the IEEE/CVF winter conference on applications of computer vision}, pages 1928--1938, 2023.

\bibitem{songsolving}
Yang Song, Liyue Shen, Lei Xing, and Stefano Ermon.
\newblock Solving inverse problems in medical imaging with score-based generative models.
\newblock In {\em International Conference on Learning Representations}.

\bibitem{sun2021coil}
Yu Sun, Jiaming Liu, Mingyang Xie, Brendt Wohlberg, and Ulugbek~S Kamilov.
\newblock Coil: Coordinate-based internal learning for tomographic imaging.
\newblock {\em IEEE Transactions on Computational Imaging}, 7:1400--1412, 2021.

\bibitem{tancik2021learned}
Matthew Tancik, Ben Mildenhall, Terrance Wang, Divi Schmidt, Pratul~P Srinivasan, Jonathan~T Barron, and Ren Ng.
\newblock Learned initializations for optimizing coordinate-based neural representations.
\newblock In {\em Proceedings of the IEEE/CVF Conference on Computer Vision and Pattern Recognition}, pages 2846--2855, 2021.

\bibitem{tancik2020fourfeat}
Matthew Tancik, Pratul~P. Srinivasan, Ben Mildenhall, Sara Fridovich-Keil, Nithin Raghavan, Utkarsh Singhal, Ravi Ramamoorthi, Jonathan~T. Barron, and Ren Ng.
\newblock Fourier features let networks learn high frequency functions in low dimensional domains.
\newblock {\em NeurIPS}, 2020.

\bibitem{van2011iterative}
Gert Van~Gompel, Katrien Van~Slambrouck, Michel Defrise, K~Joost Batenburg, Johan De~Mey, Jan Sijbers, and Johan Nuyts.
\newblock Iterative correction of beam hardening artifacts in ct.
\newblock {\em Medical physics}, 38(S1):S36--S49, 2011.

\bibitem{wu2023unsupervised}
Qing Wu, Lixuan Chen, Ce Wang, Hongjiang Wei, S~Kevin Zhou, Jingyi Yu, and Yuyao Zhang.
\newblock Unsupervised polychromatic neural representation for ct metal artifact reduction.
\newblock {\em Advances in Neural Information Processing Systems}, 36:69605--69624, 2023.

\bibitem{wu2023self}
Qing Wu, Ruimin Feng, Hongjiang Wei, Jingyi Yu, and Yuyao Zhang.
\newblock Self-supervised coordinate projection network for sparse-view computed tomography.
\newblock {\em IEEE Transactions on Computational Imaging}, 9:517--529, 2023.

\bibitem{wu2021drone}
Weiwen Wu, Dianlin Hu, Chuang Niu, Hengyong Yu, Varut Vardhanabhuti, and Ge Wang.
\newblock Drone: Dual-domain residual-based optimization network for sparse-view ct reconstruction.
\newblock {\em IEEE Transactions on Medical Imaging}, 40(11):3002--3014, 2021.

\bibitem{xie2023dense}
Wangduo Xie and Matthew~B Blaschko.
\newblock Dense transformer based enhanced coding network for unsupervised metal artifact reduction.
\newblock In {\em International Conference on Medical Image Computing and Computer-Assisted Intervention}, pages 77--86. Springer, 2023.

\bibitem{ying2019x2ct}
Xingde Ying, Heng Guo, Kai Ma, Jian Wu, Zhengxin Weng, and Yefeng Zheng.
\newblock X2ct-gan: reconstructing ct from biplanar x-rays with generative adversarial networks.
\newblock In {\em Proceedings of the IEEE/CVF conference on computer vision and pattern recognition}, pages 10619--10628, 2019.

\bibitem{yu2010fast}
Zhou Yu, Jean-Baptiste Thibault, Charles~A Bouman, Ken~D Sauer, and Jiang Hsieh.
\newblock Fast model-based x-ray ct reconstruction using spatially nonhomogeneous icd optimization.
\newblock {\em IEEE Transactions on image processing}, 20(1):161--175, 2010.

\bibitem{yuce2022structured}
Gizem Y{\"u}ce, Guillermo Ortiz-Jim{\'e}nez, Beril Besbinar, and Pascal Frossard.
\newblock A structured dictionary perspective on implicit neural representations.
\newblock In {\em Proceedings of the IEEE/CVF Conference on Computer Vision and Pattern Recognition}, pages 19228--19238, 2022.

\bibitem{zang2021intratomo}
Guangming Zang, Ramzi Idoughi, Rui Li, Peter Wonka, and Wolfgang Heidrich.
\newblock Intratomo: self-supervised learning-based tomography via sinogram synthesis and prediction.
\newblock In {\em Proceedings of the IEEE/CVF International Conference on Computer Vision}, pages 1960--1970, 2021.

\bibitem{zha2022naf}
Ruyi Zha, Yanhao Zhang, and Hongdong Li.
\newblock Naf: neural attenuation fields for sparse-view cbct reconstruction.
\newblock In {\em International Conference on Medical Image Computing and Computer-Assisted Intervention}, pages 442--452. Springer, 2022.

\bibitem{zhang2021adaptive}
Marvin Zhang, Henrik Marklund, Nikita Dhawan, Abhishek Gupta, Sergey Levine, and Chelsea Finn.
\newblock Adaptive risk minimization: Learning to adapt to domain shift.
\newblock {\em Advances in Neural Information Processing Systems}, 34:23664--23678, 2021.

\bibitem{zhang2018sparse}
Zhicheng Zhang, Xiaokun Liang, Xu Dong, Yaoqin Xie, and Guohua Cao.
\newblock A sparse-view ct reconstruction method based on combination of densenet and deconvolution.
\newblock {\em IEEE transactions on medical imaging}, 37(6):1407--1417, 2018.

\end{thebibliography}
}
\onecolumn
\section*{\LARGE Appendix}
\vspace{10pt}
\setcounter{section}{0}
\section{Visual Comparison on Ellipse Material Dataset}

\cref{fig:ellipse_recon_20} and \cref{fig:ellipse_recon_40} show the reconstruction results of different methods on the Ellipse Material Dataset. The PSNR/SSIM values of AC-IND(Ours) are marked in \textcolor{darkyellow}{yellow}. The PSNR/SSIM values of AC-IND\textsuperscript{+}(Ours\textsuperscript{+}) are marked in \textcolor{red}{red}.  Please zoom in to see more details.
\setcounter{figure}{0}
\begin{figure*}[h]
    \centering
    \includegraphics[width=\textwidth]{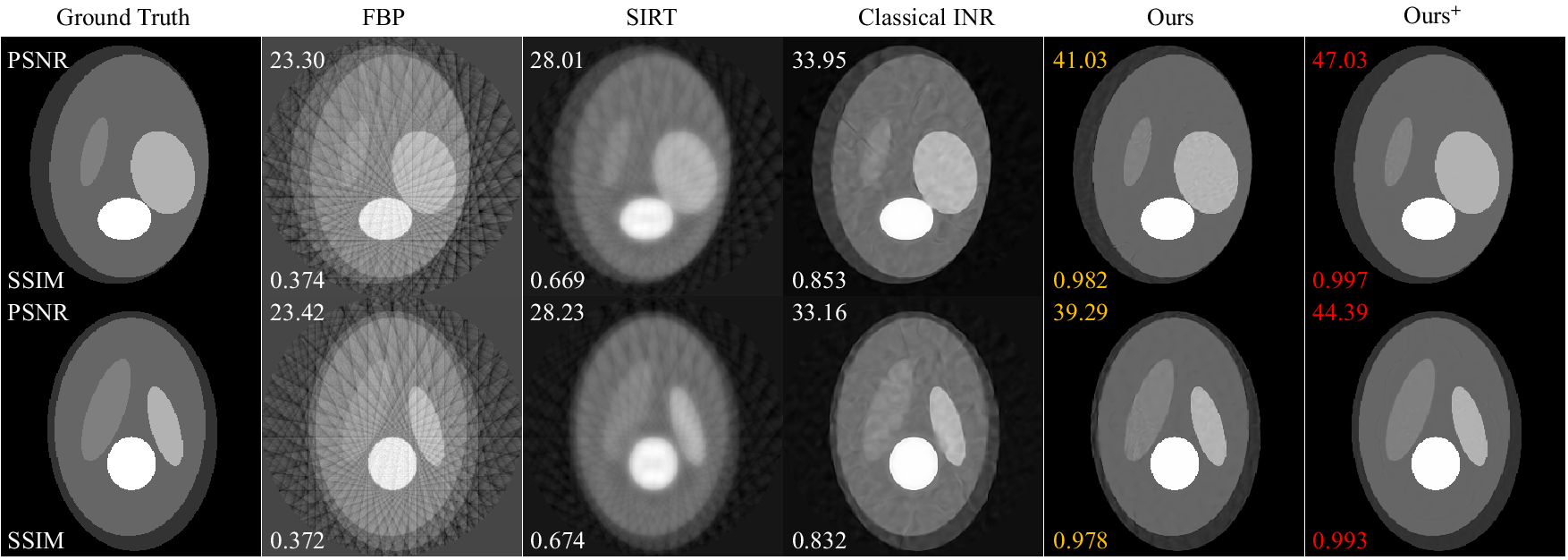} 
    \caption{Comparison of different reconstruction under \textbf{20 views} setting.}
    \label{fig:ellipse_recon_20}
\end{figure*}

\begin{figure*}[h]
    \centering
    \includegraphics[width=\textwidth]{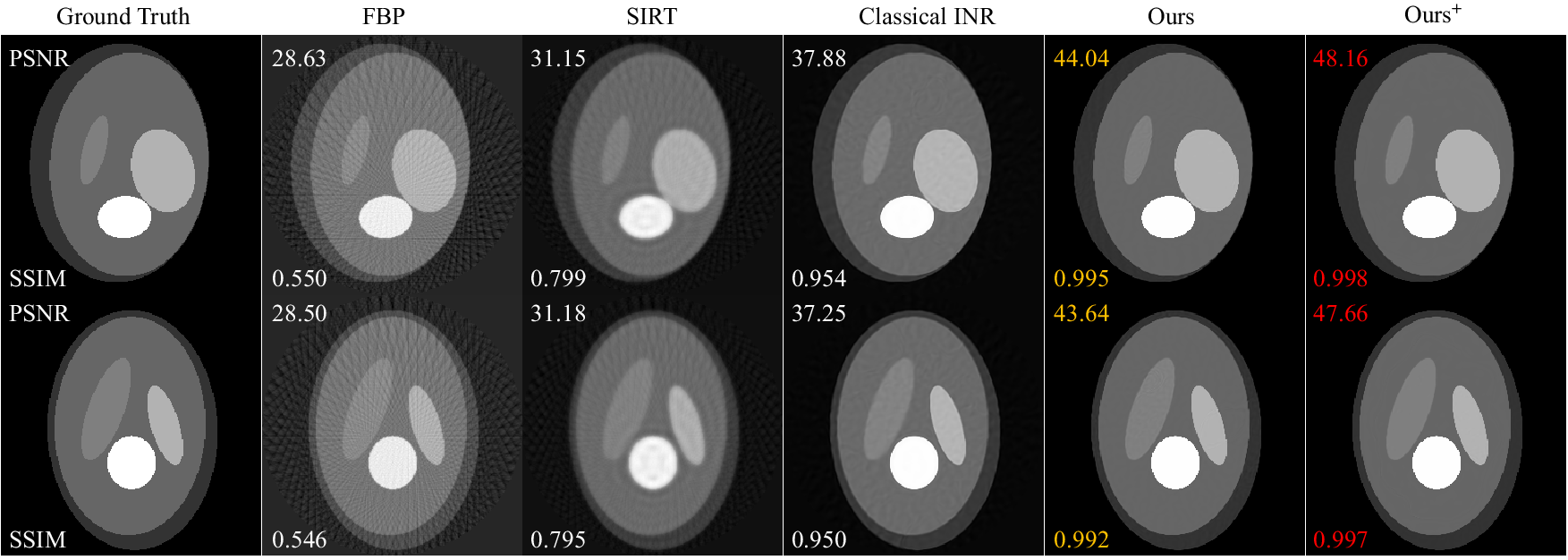} 
    \caption{Comparison of different reconstruction under \textbf{40 views} setting.}
    \label{fig:ellipse_recon_40}
\end{figure*}
\section{Visual Comparison on Walnut Slice Dataset}

\begin{figure*}[htbp]
    \centering
    \includegraphics[width=\textwidth]{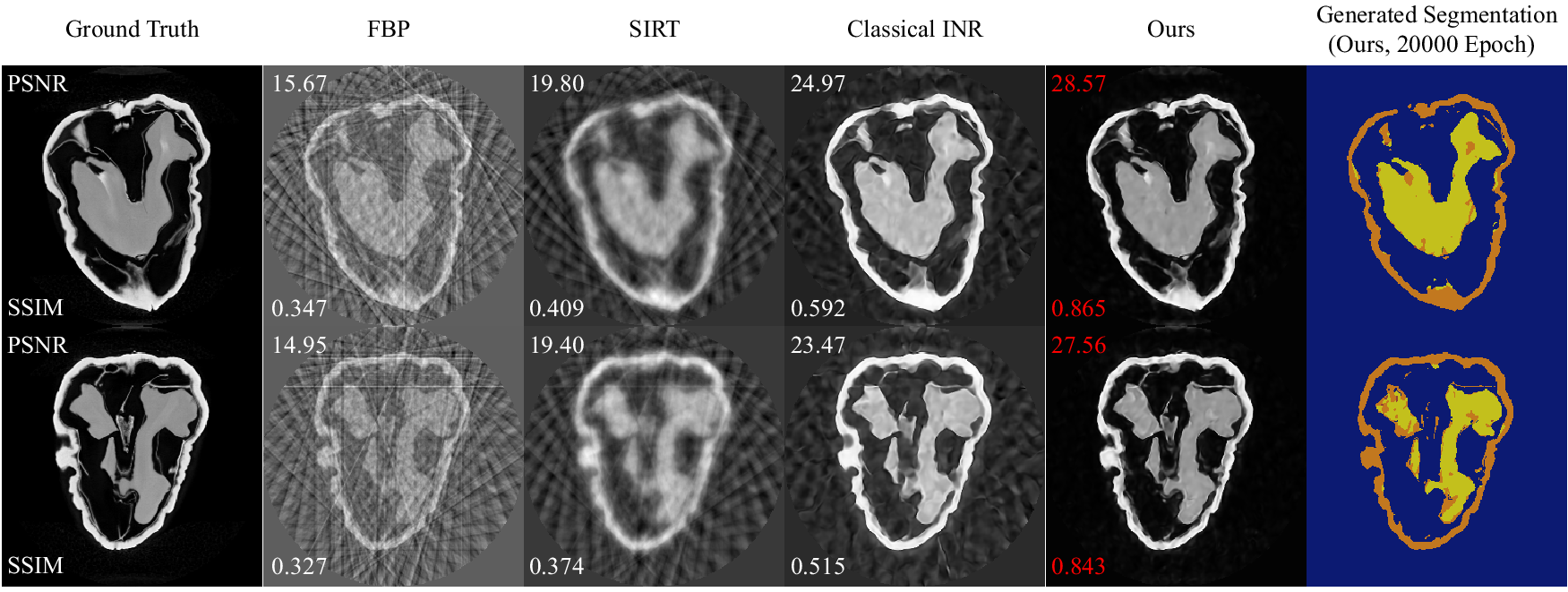} 
    \caption{Comparison of different reconstruction under \textbf{20 views} setting, and the segmentation map generated by our method.}
    \label{fig:walnut_recon_20}
\end{figure*}

\begin{figure*}[htbp]
    \centering
    \includegraphics[width=\textwidth]{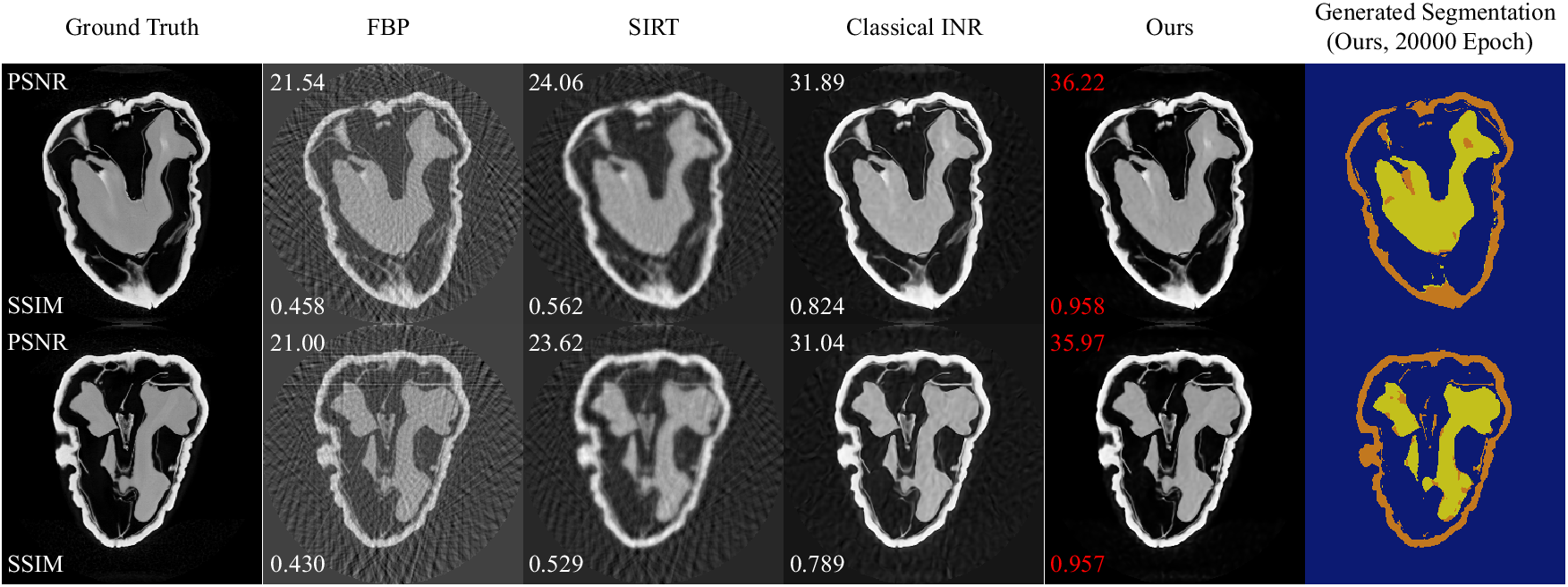} 
    \caption{Comparison of different reconstruction under \textbf{40 views} setting, and the segmentation map generated by our method.}
    \label{fig:walnut_recon_40}
\end{figure*}

\begin{figure*}[htbp]
    \centering
    \includegraphics[width=\textwidth]{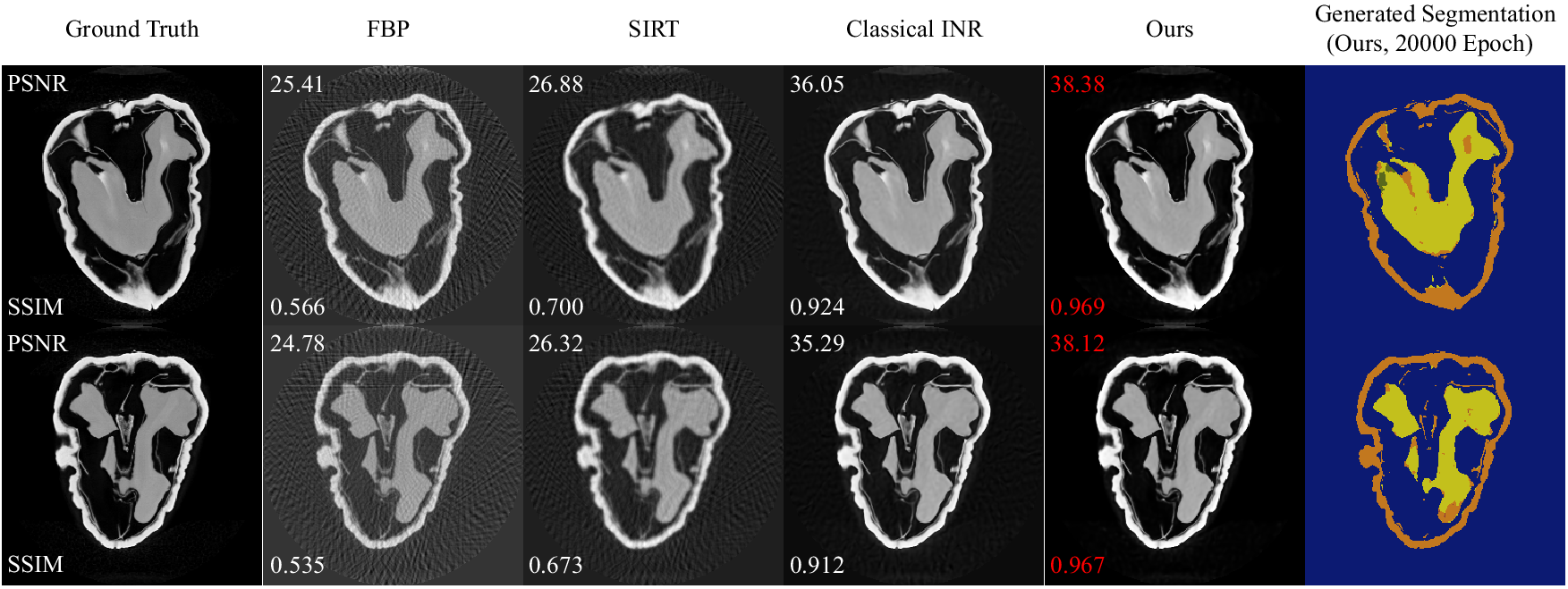} 
    \caption{Comparison of different reconstruction under \textbf{60 views} setting, and the segmentation map generated by our method.}
    \label{fig:walnut_recon_60}
\end{figure*}

\cref{fig:walnut_recon_20}, \cref{fig:walnut_recon_40} and \cref{fig:walnut_recon_60} show the reconstruction results of different methods on the Walnut Slice Dataset. In addition, the semantic segmentation maps automatically generated by AC-IND are also shown in \cref{fig:walnut_recon_20}, \cref{fig:walnut_recon_40} and \cref{fig:walnut_recon_60}.  The PSNR/SSIM values of AC-IND(Ours) are marked in \textcolor{red}{red}. Please zoom in to see more details.

\end{document}